# Guest Editorial: A Critical Review of Recent Progress on Negative Capacitance Field-Effect Transistors

Muhammad A. Alam [1], Mengwei Si [2], and Peide D. Ye [3]

School of Electrical and Computer Engineering, Purdue University, West Lafayette, IN 47907, USA

Corresponding authors: Muhammad A. Alam and Peide D. Ye

E-mail: [1] alam@purdue.edu, [2] msi@purdue.edu, [3] yep@purdue.edu

The astounding progress of electronics in the 20th century sometimes obscures the dramatic story of repeated reinvention of the underlying device technology. The reinventions were catalyzed by the limits of power dissipation and self-heating for the corresponding device technologies. In the 1950s, when the vacuum-tubes reached the power dissipation limits, the more power efficient bipolar transistors took over. In the 1980s, bipolar transistors were replaced by even more power efficient technology based on complementary metal-oxide-semiconductor (CMOS) field-effect transistors (FET). As power consumption, self-heating and scaling considerations threaten the scaling of CMOS at the twilight of Moore's law, it is not surprising that researchers are once again looking for a more scalable and energy-efficient replacement, such as tunnel FET[1], Nano-Electro-Mechanical (NEM)-FET[2], spin-FET[3], phase-FET[4], etc. The negative capacitance field-effect transistor (NC-FET) proposed by Salahuddin and Datta[5] is a recent entry to the list.

The so-called Boltzmann tyranny defines the fundamental thermionic limit of the subthreshold slope (SS) of a MOSFET at 60 mV/dec at room temperature, and therefore precludes lowering of the supply voltage and overall power consumption. As shown in Fig. 1(a), an NC-FET adds a thin-layer of ferroelectric (FE) material to the existing gate oxide of a MOSFET. The theory suggests that the consequence of this "trivial" change can be dramatic with complete disappearance of ferroelectric hysteresis ($\Delta V$), Fig. 1(b). The internal voltage at the FE-oxide interface would be *larger* than the gate voltage, so that the SS will reduce below the Boltzmann limit of 60 mV/dec at room temperature, as shown in Fig. 1(c). As a result, the on-current ($I_{on}$) would be reached at a lower supply voltage ($V_{DD}$) and the power consumption would be reduced significantly. Moreover, Fig. 1(d) shows that unlike a traditional MOSFET, the threshold voltage ($V_{th}$) would actually *increase* as $V_{DD}$ increases, making transistor scaling easier. The elegant simplicity of the device concept and the urgent need for a new "transistor" at the twilight of the Moore's law have inspired many researchers in industry and academia to explore the physics and technology of the NC-FET and since 2008, hundreds of papers have been published.

Despite the simplicity of the original NC-FET theory[5], the experimental data accumulated over the years (Fig. 1(e)) show a relatively broad scatter. This level of scatter is not unexpected for a fundamentally new class of transistor. However, in the context of the frantic pace of activities, scatter in the published data, challenges of characterization, and emergence of a diversity of models used to interpret the results, some researchers have asked thoughtful and interesting questions regarding the physics and viability of the device technology which are summarized in the following discussion.

The questions relate to a basic issue about the polarization of the thin ferroelectric layer shown in Fig. 1(a). In an NC-FET, the series addition of a sufficiently large positive (i.e. gate or depletion) capacitor is expected to stabilize the FE in the zero polarization state. In the original NC-FET theory[5], the NC effect (also called "quasi-static NC") is realized without polarization switching. In contrast, the "transient NC" effect requires and is associated with real polarization switching. Theoretically, the zero polarization state can be interpreted either by a single-domain or a multi-domain model. In the single domain approximation, each unit cell of the stabilized FE individually has zero polarization because the atoms are symmetrically placed within the cell and the charge centroids are co-located so that it resembles a normal dielectric under zero electric field, but with an enhanced capacitance. In a multi-domain approximation, the zero polarization is achieved by having equal number of up vs. down polarized unit cells. Traditionally, many phenomena of ferroelectricity have been explained in terms of a multi-domain approximation, while the single domain approximation is essential to the operation of an NC-FET. To distinguish between single vs. multi-domain interpretations, the confusion arises because the underlying issues are often phrased in terms of a variety of related or unrelated questions. For clarity, these questions must be addressed separately and the interpretations must be self-consistent, i.e. a single model must interpret all the relevant experiments. The goals of this article are (a) to organize and compare the results from various experiments and modeling efforts published to-date, (b) to use the information gathered to answer a set of important questions in the field, and (c) to suggest a protocol for reporting NC-FET experiments. In this rapidly evolving field, we cannot offer conclusions, but simply provide some starting points for a coherent discussion.

In essence, there are three types of questions regarding an NC-FET.

1. *Can a capacitor be negative? Is there any device that unambiguously demonstrated negative capacitance?*

   The answer is yes. The negative capacitance associated with a micro-electro-mechanical (MEM) switch can be unambiguously stabilized at any position within the unstable region, demonstrating the existence and utility of the negative capacitance. A number of experimental results and theoretical calculations support this concept[2,6]. Although impractical as a MOSFET replacement, a MEM-switch offers conceptual clarity regarding negative capacitance, because its operation is characterized by a single order parameter related to the air gap.

2. *Given the domain dynamics, can a FE-based capacitor or FET show negative capacitance? Are the reported transient and steady-state experiments conclusive?*

   Over the years, four types of experiments shown in Fig. 2 have addressed this question. The *stacked capacitor-based* approach measures the transient and steady-state responses of a dielectric (DE)/ferroelectric series capacitor to see if the internal voltage is indeed amplified. When the internal node is accessible, the node voltage can be measured directly. Otherwise, one compares the total capacitance with the known capacitance of the dielectric, $C_{DE}$. **One thing to be careful about** is that there may be a fundamental difference between a capacitor with FE/DE heterostructure vs. a DE capacitor and a FE capacitor connected in series[7,8]. It has been reported that even the measurement setup affects the results significantly and voltage meter with sufficiently high impedance is needed in the measurement of internal $V_{INT}$[9].

In the **small-signal measurement** (Fig. 2(a)), the total capacitance of the stack (at $V_G = 0$) is reported to be larger than the capacitance of the dielectric layer[10–13], suggesting the validity of the single-domain approximation. Such DC enhancement was not observed in multi-domain ferroelectric HZO externally connected to a commercial DE capacitor[7], or in a FE/DE stack[8,14]. The small-signal measurement is considered as a quasi-static measurement without triggering the polarization switching[7]. To explain the apparent discrepancy, some researchers have suggested that the metal interlayer (needed to measure the internal node voltage) fundamentally alters the FE-polarization[8,9]. A precise mathematical formulation of the essential difference between the two structures is still being formulated.

In the **transient RC measurements** (Fig. 2(b)), a voltage drop across the ferroelectric capacitor is observed when applying a voltage pulse[15–18]. Initially, the phenomenon was interpreted by a single-domain Landau-Khalatnokov (L-K) model with renormalized parameters and was taken as an unambiguous proof of the existence of a negative capacitance effect in the ferroelectric insulator[15]. Recently, other groups have argued that multi-domain variants, such as the Kolmogorov–Avrami–Ishibashi (KAI) model[19] or Preisach-Miller models[20–22] can also explain the experimental observations. They attributed the observation to the delay in domain flipping associated with polarization switching and discharging of the dielectric components[19,21,23,24]. This alternate explanation suggests that the transient experiments may not be able to conclusively distinguish between single and multi-domain dynamics.

A third type of measurement is the **ramp voltage pulse** measurement of FE and DE capacitor in series as shown in Fig. 2(c). A differential voltage amplification on DE capacitor by the FE capacitor was observed[25]. Once again, the result can be explained either by single domain or multi-domain switching model[22].

Although the small signal, pulse, and ramp voltage experiments have not produced definitive conclusion, they have highlighted the need to distinguish between samples with and without internal nodes, thin vs. thick ferroelectrics, one vs. two dimensional analysis, and the importance of leakage current in interpreting the diversity of the results reported to date[8].

The second class of experiments (Fig. 2(d)) involve fabricating an NC-FET and directly measuring its subthreshold slope, on-current, drain-induced barrier lowering (DIBL), and output conductance as shown in Fig. 1(b)-(d). The scatter in the NC-FET data shown in Fig. 1(e) could be understood to result from considerations discussed next.

**Another pitfall is to confuse NC-FETs with ferroelectric FETs (Fe-FETs).** One must not confuse NC-FETs with Fe-FETs; they are structurally identical but functionally distinct. A Fe-FET has hysteretic I-V characteristics, but an NC-FET does not, see Fig. 1(b). In an NC-FET, the total gate capacitance is positive, which means that the negative capacitance state of the ferroelectric insulator is stabilized in a single state according to the quasi-static NC model.[5] In a Fe-FET, the total gate capacitance is negative (if we use the NC concept to understand ferroelectric switch), so that the transistor switches between two states with the corresponding hysteresis in the transfer characteristics as highlighted in Fig. 1(b). In addition, if the ferroelectric polarization switching in a Fe-FET happens in the "subthreshold" region, a SS with deep sub-60 mV/dec at room temperature may be observed. However, the deep sub-60 mV/dec in a Fe-FET comes from the ferroelectric polarization switching instead of the negative capacitance effect. Although a NC-FET is fundamentally different from a Fe-FET, a Fe-FET-based logic switch with hysteresis window less than half the operating voltage may still offer higher on-current and lower off-current[8]. Ultimately its adoption as a logic switch[9] will depend on the variability of the hysteresis window and the fundamental speed of (single or multiple) domain switching.

In the literature, both the Fe-FET (sometimes interpreted as an unstabilized NC-FET) and quasi-static NC-FET have been studied and reported. The first experimental reports explored the question of "negative capacitance" and steep-switching associated with Si Fe-FETs that used a thick P(VDF-TrFE) polymer as the ferroelectric insulator[26–28]. The discovery of ferroelectric $HfO_2$ was an exciting advance, because it enabled CMOS compatible processing of a ferroelectric-gated MOSFET[29,30]. Quasi-static or stabilized hysteresis-free Si NC-FETs with sub-60 mV/dec SS at room temperature with ferroelectric hafnium zirconium oxide (HZO) as the ferroelectric gate insulator have been reported beginning in 2014[31–34]. After these works, Si NC-FETs were studied with various gate stacks and structures, which fall into the category of either Fe-FET (unstabilized NC-FET)[35–40] or steep-slope hysteresis-free NC-FET[41–47] with minimum SS down to ~40 mV/dec at room temperature.

NC-FETs with alternative channel materials have also been reported. For example, NC-FETs with a Ge channel were demonstrated[48–54]. Steep subthreshold-slope and nearly hysteresis-free performance have been observed[50–53]. The first reported 2D NC-FET applied P(VDF-TrFE) polymer as ferroelectric insulator and $MoS_2$ as the channel material, but the fabricated device was unstable and double-sweep transfer characteristics were not measured[55]. In 2017, 2D $MoS_2$ steep-slope and hysteresis-free NC-FETs were demonstrated by careful capacitance matching design[56–58]. Unstabilized NC-FETs with a 2D $MoS_2$ channel were also reported. Although they achieved a sub-60 mV/dec SS at room temperature, they featured a counterclockwise hysteresis[59–62]. NC-FETs using other low dimensional materials such as carbon nanotubes[63] and $WSe_2$[64] have also been reported. Finally, an NC-FET with a III-V semiconductor as the channel was also demonstrated, but hysteresis-free and sub-60 mV/dec SS haven't been achieved simultaneously[65].

In short, a sweep-voltage and sweep-frequency independent, hysteresis-free I-V characteristic is an essential pre-requisite for definitive proof of the NC-FET operation. Figure 1(e) shows that despite a large number of NC-FET reports, only a minority of the experiments satisfy the two criteria of being hysteresis free and exhibiting sub-60 mV/dec SS simultaneously.

**Most importantly, experiments must be interpreted self-consistently.** A single-domain L-K theory anticipates simultaneous occurrence of reduced SS, negative DIBL, negative drain resistance (NDR)[5,66,67] and noise-suppression of the drain-current as shown in Fig. 1(c) and (d) and demonstrated experimentally in Refs. [57] and [68] , for example. Thus, if multiple groups were to report these features associated with a single device, it would support the existence of NC-FET operation.

It is important to understand that "multi-domain models" derived from Grinzburg-Landau theory have so far not been able to explain the observed NDR, negative DIBL, and *hysteresis-free* sub-60 mV/dec SS directly and self-consistently. Rather, the specialized multi-domain models (e.g. KAI[19], Miller[20–22,69], and/or Modified Miller[70]) interpret the steady state response by suggesting that all steady-state measurements are in-fact time-dependent, defined by the sweep rate of measurement. Thus, they interpret the "DC" subthreshold slope as a consequence of time-dependent phase-lag, associated with ferroelectric polarization switching[9,21,70–73]. Also, some models attempt to explain the hysteresis-free operation and NDR by invoking non-ideal charge trapping to compensate the counterclockwise hysteresis[21,74]. Unfortunately, the charge trapping leads to substantially different forward and reverse subthreshold sweeps[74] and cannot explain (essentially) hysteresis-free operation seen in many experiments, as in the bottom left corner of Fig. 1e. A self-consistent explanation of the observed features remains an important goal for "multi-domain" theory of NC-FET operation.

3. *Even if a FE-DE can be stabilized in the NC state, are the dimensions suitable for ultra-scaled transistors beyond the 5 nm technology node? Would it switch fast enough? Given the unique physics of the gate stack, would the technology be reliable and immune from gate dielectric breakdown, negative bias temperature instability, hot carrier degradation, and other perspectives?*

The scaling questions are device specific. For example, several groups have reported that the parasitic gate-drain capacitance of a FinFET actually improves capacitance matching and reduces subthreshold slope[66,67,75]. On the other hand, the quantum capacitance of ultra-thin body transistors may negate some of the improvement. Recently, researchers from Global Foundries have reported integrating doped hafnium oxide ferroelectric layers into state-of-the-art 14nm Si FinFET technology and demonstrated that 101 stage ring oscillators show improved SS and actually reduce the active power consumption of the circuits[43].

The frequency dependence is another question of interest. Quasi-static NC-FET models argue that transistor operation does not require domain switching[5,43], so that FE switching speed may not be relevant for NC-FET operation. An in-depth recent analysis[76] however concludes that an NC-FET will always switch slower than the corresponding Fe-FET. This limit reflects the fact that while the amount of polarization switching necessary for an NC-FET is substantially smaller than that of a Fe-FET, the internal field in the NC-FET is also substantially smaller than that of the Fe-FET. Therefore, one can view Fe-FET switching speed (see Fig. 1(f)) as the upper limit of NC-FET switching. A recent experiment has reported 3.6 ns single pulse response and 100 ps multi-pulse response of a HfO$_2$-based ferroelectric switch[77], suggesting the possibility of achieving near GHz operation. Single ultra-short pulse measurement could be just limited by obtaining sufficient inversion charges to support FE switching. Finally, a report based on optical characterization of polarization switching[78] and recent experiments by Global Foundries[43,79] indicate tens-of-GHz switching may be possible. Fig. 1(f) summarizes the representative reports in terms of ferroelectric switching speed and compares NC-FETs to the current Si transistor technology[80]. The unification of various characterization methodologies and quantification of the damping coefficient in the L-K equation are essential for future progress regarding this topic. Unless new data shows otherwise, one may be cautiously optimistic regarding the switching speed of these transistors.

Finally, reliability issues place several important constraints on device operation[81–84]. The voltage application at the dielectric node of the gate-stack suggests that dielectric breakdown considerations would restrict NC-FET operation at the same on-current with the same interface field but at a reduced operating voltage. It has been suggested that a V-shaped field profile in the gate-stack[9] would lead to bias-temperature instability (BTI) issues related to collection of the tunneling, soft-breakdown, and hot carrier injection (HCI)-induced current at the dielectric/ferroelectric interface[85,86]. Fortunately, the interface of defect generation, parasitic gate-drain capacitance, and negative capacitance is likely to suppress the Negative-bias temperature instability (NBTI) degradation – the most important reliability concern for modern MOSFETs[87]. Also, since the HZO transition temperature is sufficiently high, self-heating induced changes of the Landau-coefficients may not be an important issue. Transistor reliability is fundamentally important and establishing NC-FET reliability would be an important goal for future research.

The discussion above related to these three questions highlights the fact that a fragmented approach (that only emphasizes the reduction of subthreshold slope) has resulted in a confusing mix of results in the field. The NC-FET concept must be self-consistently validated by (a) combination of dielectric and ferroelectric thicknesses, (b) a broad set of sweep ranges and rates, (c) comprehensive reports of transfer, output, and noise characteristics, demonstrating hysteresis free operation, negative DIBL, negative drain resistance, and suppression of 1/f noise. It is important to report frequency response of an isolated NC-FET to establish a lower limit of operation of these transistors. Reliability studies of NC-FETs, i.e., stability of charge accumulation and threshold voltage, voltage acceleration and ferroelectric dielectric stack breakdown, are urgently needed. Obviously, as transistor technology scales below 5 nm node, the critical device dimensions are extremely small and the questions of integrating sufficiently-thick FE-layers into a gate stack as well as ensuring high speed and reliability would become increasing important questions.

The NC-FET concept provides a unifying perspective to a broad range of device phenomena collectively known as Landau switches and it allows arbitrary tailoring of the energy landscape[99]. Although the validity of quasi-static NC is still being debated, the concept of NC – if conclusively demonstrated – will have broad implications for device physics. Indeed, its conceptual demonstration would open up a broad class of applications including electro-chemical sensing and MEMS-based actuation[88]. In this regard, the experience of thick ferroelectric films should inform, but not constraint future research in the field. The $HfO_2$-based ferroelectric films are relatively new to the material/device communities, therefore their properties may be substantially different from traditional ferroelectric materials. The domain dynamics of such a constrained thin film is indeed not known. Given the urgency of finding a new low-power switch, NC-FET research justifiably merits the broad attention and in-depth analysis it has received from the device physics community over the last decade. See supplemental materials for a summary of all the representative models.

The authors acknowledge the valuable discussions with Supriyo Datta, M.S. Lundstrom, K. Karda, S. Salahuddin, Suman Datta, A.I. Khan, K.K. Ng, S.K. Gupta, J. van Houdt, and many other experts in the field. The work was supported in part by ASCENT, one of six centers in JUMP, a semiconductor Research Corporation (SRC) program sponsored by DARPA.

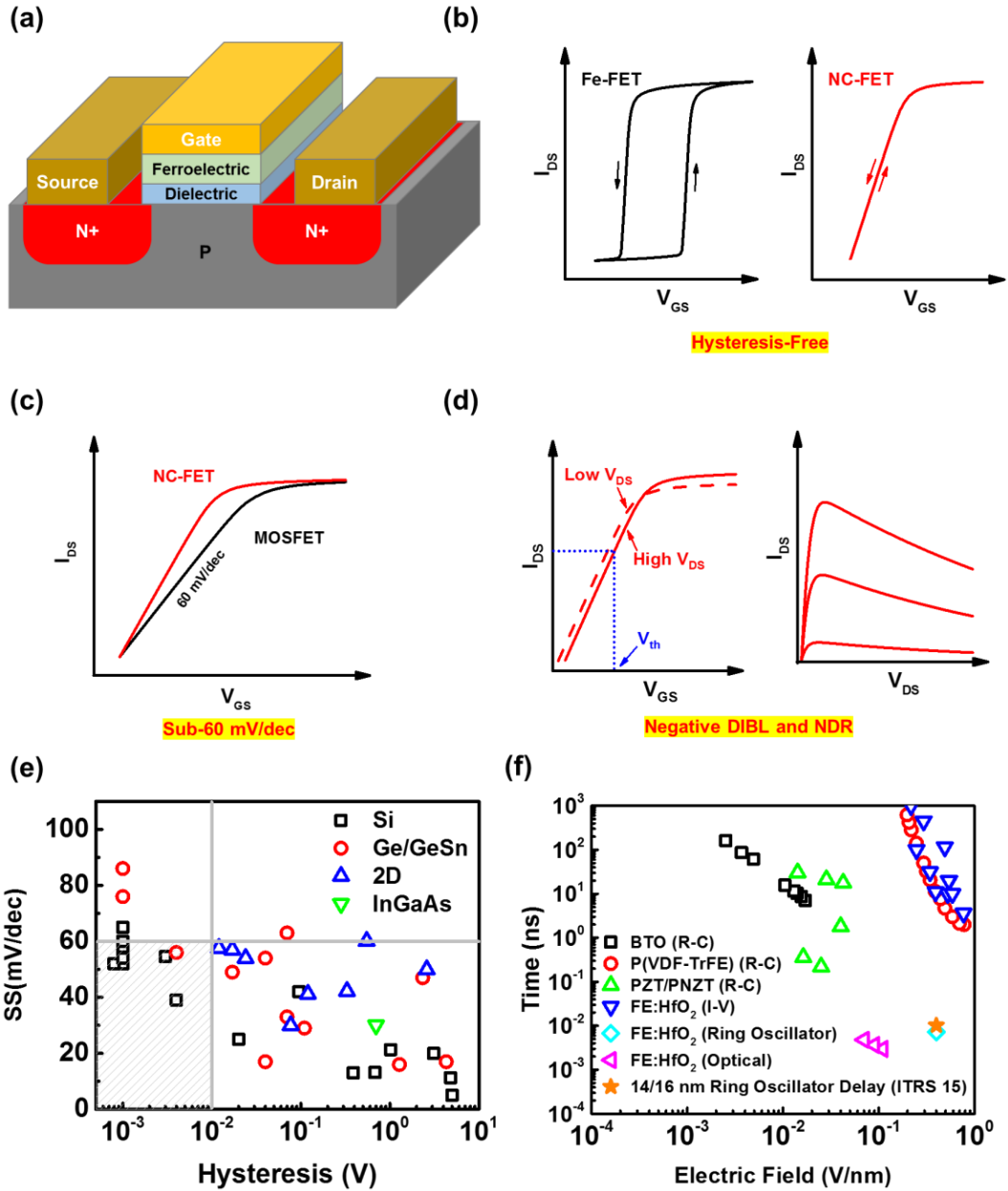

Figure 1. (a) Schematic image of an NC-FET with ferroelectric and conventional dielectric as the gate stack. (b) The fundamental difference in transfer characteristics of a Fe-FET versus an NC-FET which has an anti-clockwise hysteresis or zero-hysteresis respectively. (c) Expected steep-slope less the 60 mV/dec at room temperature for a NC-FET. (d) Expected negative DIBL and negative drain resistance for a NC-FET. (e) Summary of reported representative data in literature in terms of SS versus hysteresis in transfer characteristics (Si[26,28,33-36,41-45,47], Ge/GeSn[48-53], 2D[56-60], InGaAs[65]). SS is plotted as the larger SS in forward and reverse gate sweeps and only when both are available. Data without explicitly reported hysteresis are plotted with 1 mV hysteresis. (f) Summary of reported switch times of representative ferroelectric films versus electric field by different characterization methods in the literature. (BTO (R-C)[89], PZT/PNZT (R-C)[90-93], P(VDF-TrFE) (R-C)[94], FE:HfO$_2$ (I-V)[77,95-98], FE:HfO$_2$ (Ring Oscillator)[43,79], FE:HfO$_2$ (Optical)[78])

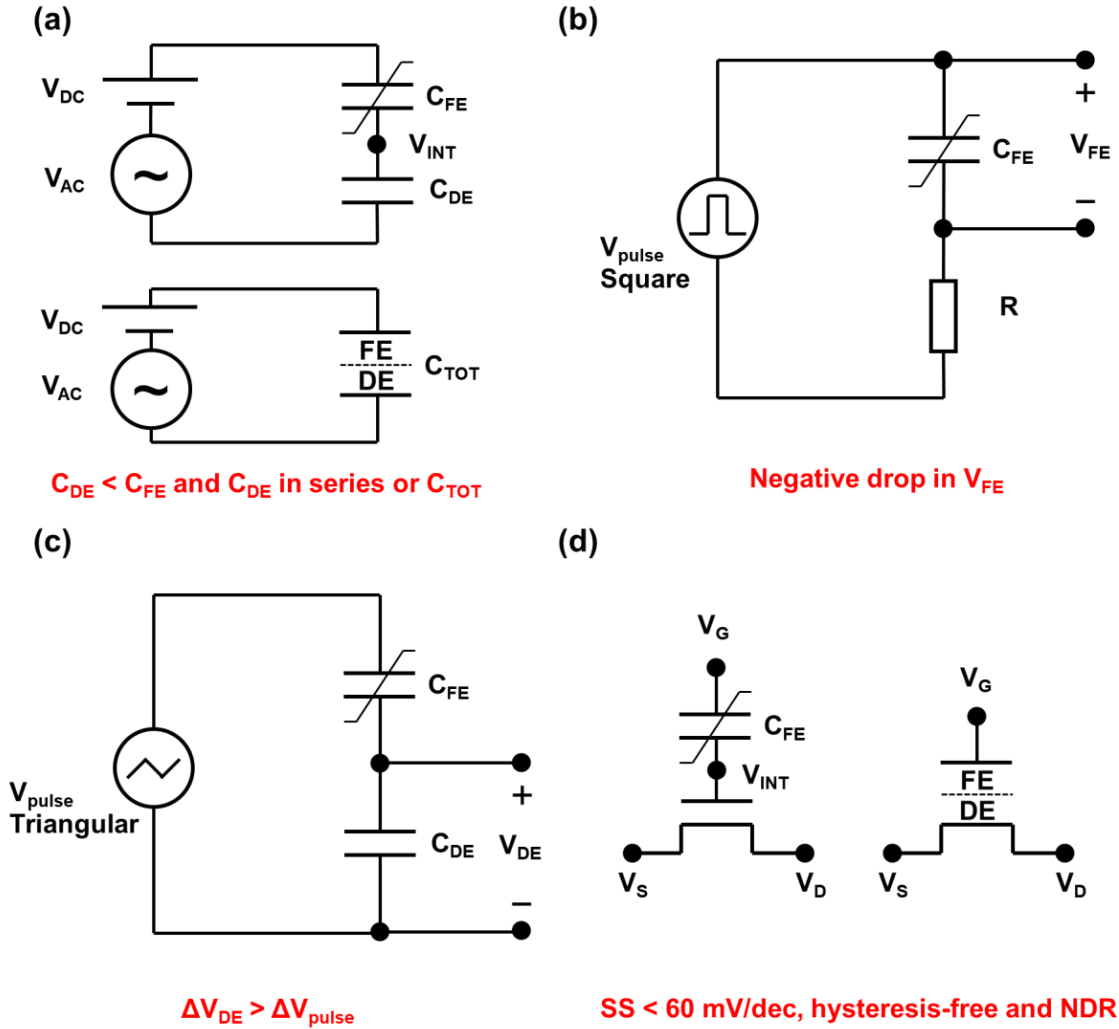

Figure 2. Four types of experiments have been used to characterize the negative capacitance effect. (a) Two configurations for the small signal measurement: (i) The internal metallic node separating $C_{DE}$ and $C_{FE}$ capacitances is used to measure the voltage/capacitances, and (ii) total capacitance of FE/DE stack is measured and compared to $C_{DE}$. (b) Transient RC measurement: Unlike typical RC decay, the voltage across the ferroelectric capacitor may actually increase. (c) Ramp pulse measurement: Voltage change across the $C_{DE}$ ($\Delta V_{DE}$) could be larger than voltage change in $V_{pulse}$ ($\Delta V_{pulse}$) as a consequence of voltage amplification. (d) Transistor measurement: Sub-60 mV/dec subthreshold slope at room temperature, hysteresis-free and negative differential resistance are signatures of NC-FET.

# Reference

[1] A.M. Ionescu and H. Riel, Nature **479**, 329 (2011).

[2] C. Goldsmith, T.-H. Lin, B. Powers, W.-R. Wu, and B. Norvell, in *IEEE MTT S Int. Microw. Symp. Dig.* (1995), p. 91.

[3] S. Datta and B. Das, Appl. Phys. Lett. **56**, 665 (1990).

[4] N. Shukla, A. V Thathachary, A. Agrawal, H. Paik, A. Aziz, D.G. Schlom, S.K. Gupta, R. Engel-Herbert, and S. Datta, Nat. Commun. **6**, 7812 (2015).

[5] S. Salahuddin and S. Datta, Nano Lett. **8**, 405 (2008).

[6] H. Kam and T.-J.K. Liu, IEEE Trans. Electron Devices **56**, 3072 (2009).

[7] Z. Liu, M.A. Bhuiyan, and T.P. Ma, in *IEEE Intl. Electron Devices Meet.* (2018), pp. 711–714.

[8] M. Si, X. Lyu, and P.D. Ye, arXiv:1812.05260 (2018).

[9] X. Li and A. Toriumi, in *IEEE Intl. Electron Devices Meet.* (2018), pp. 715–718.

[10] A.I. Khan, D. Bhowmik, P. Yu, S.J. Kim, X. Pan, R. Ramesh, and S. Salahuddin, Appl. Phys. Lett. **99**, 113501 (2011).

[11] W. Gao, A. Khan, X. Marti, C. Nelson, C. Serrao, J. Ravichandran, R. Ramesh, and S. Salahuddin, Nano Lett. **14**, 5814 (2014).

[12] D.J.R. Appleby, N.K. Ponon, K.S.K. Kwa, B. Zou, P.K. Petrov, T. Wang, N.M. Alford, and A. O'Neill, Nano Lett. **14**, 3864 (2014).

[13] P. Zubko, J.C. Wojdeł, M. Hadjimichael, S. Fernandez-Pena, A. Sené, I. Luk'yanchuk, J.-M. Triscone, and J. Íñiguez, Nature **534**, 524 (2016).

[14] M. Hoffmann, B. Max, T. Mittmann, U. Schroeder, S. Slesazeck, and T. Mikolajick, in *IEEE Intl. Electron Devices Meet.* (2018), pp. 727–730.

[15] A.I. Khan, K. Chatterjee, B. Wang, S. Drapcho, L. You, C. Serrao, S.R. Bakaul, R. Ramesh, and S. Salahuddin, Nat. Mater. **14**, 182 (2015).

[16] M. Hoffmann, M. Pešić, K. Chatterjee, A.I. Khan, S. Salahuddin, S. Slesazeck, U. Schroeder, and T. Mikolajick, Adv. Funct. Mater. **26**, 8643 (2016).

[17] M. Kobayashi, N. Ueyama, K. Jang, and T. Hiramoto, in *IEEE Intl. Electron Devices Meet.* (2016), pp. 314–317.

[18] P. Sharma, J. Zhang, K. Ni, and S. Datta, IEEE Electron Device Lett. **39**, 272 (2018).

[19] Y.J. Kim, H.W. Park, S.D. Hyun, H.J. Kim, K. Do Kim, Y.H. Lee, T. Moon, Y. Bin Lee, M.H. Park, and C.S. Hwang, Nano Lett. **17**, 7796 (2017).

[20] M. Hoffmann, A.I. Khan, C. Serrao, Z. Lu, S. Salahuddin, M. Pešić, S. Slesazeck, U. Schroeder, and T. Mikolajick, J. Appl. Phys. **123**, 184101 (2018).

[21] B. Obradovic, T. Rakshit, R. Hatcher, J.A. Kittl, and M.S. Rodder, in *VLSI Tech. Dig.* (2018), pp. 51–52.

[22] A.K. Saha, S. Datta, and S.K. Gupta, J. Appl. Phys. **123**, 105102 (2018).

[23] S.C. Chang, U.E. Avci, D.E. Nikonov, S. Manipatruni, and I.A. Young, Phys. Rev. Appl. **9**, 14010 (2018).

[24] K. Ng, S.J. Hillenius, and A. Gruverman, Solid State Commun. **265**, 12 (2017).

[25] A.I. Khan, M. Hoffmann, K. Chatterjee, Z. Lu, R. Xu, C. Serrao, S. Smith, L.W. Martin, C.C. Hu, R. Ramesh, and S. Salahuddin, Appl. Phys. Lett. **111**, 253501 (2017).

[26] G.A. Salvatore, D. Bouvet, and A.M. Ionescu, in *IEEE Intl. Electron Devices Meet.* (2008), pp. 479–481.

[27] A. Rusu, G.A. Salvatore, D. Jiménez, and A.M. Ionescu, in *IEEE Intl. Electron Devices Meet.* (2010), pp. 395–398.

[28] J. Jo, W.Y. Choi, J.D. Park, J.W. Shim, H.Y. Yu, and C. Shin, Nano Lett. **15**, 4553 (2015).

[29] T.S. Böescke, J. Müller, D. Bräuhaus, U. Schröder, and U. Böttger, in *IEEE Intl. Electron Devices Meet.* (2011), pp. 547–550.

[30] J. Muller, T.S. Boscke, U. Schroder, S. Mueller, D. Brauhaus, U. Bottger, L. Frey, and T. Mikolajick, Nano Lett **12**, 4318 (2012).

[31] C.H. Cheng and A. Chin, IEEE Electron Device Lett. **35**, 274 (2014).

[32] M.H. Lee, Y. Wei, K. Chu, J. Huang, C. Chen, C. Cheng, M. Chen, H. Lee, Y. Chen, L. Lee, and M. Tsai, IEEE Electron Device Lett. **36**, 294 (2015).

[33] M.H. Lee, P. Chen, C. Liu, K. Chu, C. Cheng, M. Xie, S. Liu, J. Lee, M. Liao, M. Tang, K. Li, and M. Chen, in *IEEE Intl. Electron Devices Meet.* (2015), pp. 616–619.

[34] K.-S. Li, P.-G. Chen, T.-Y. Lai, C.-H. Lin, C.-C. Cheng, C.-C. Chen, Y.-J. Wei, Y.-F. Hou, M.-H. Liao, M.-H. Lee, M.-C. Chen, J.-M. Sheih, W.-K. Yeh, F.-L. Yang, S. Salahuddin, and C. Hu, in *IEEE Intl. Electron Devices Meet.* (2015), pp. 620–623.

[35] A.I. Khan, K. Chatterjee, J.P. Duarte, Z. Lu, A. Sachid, S. Khandelwal, R. Ramesh, C. Hu, and S. Salahuddin, IEEE Electron Device Lett. **37**, 111 (2016).

[36] E. Ko, J.W. Lee, and C. Shin, IEEE Electron Device Lett. **38**, 418 (2017).

[37] P. Sharma, K. Tapily, A.K. Saha, J. Zhang, A. Shaughnessy, A. Aziz, G.L. Snider, S. Gupta, R.D. Clark, and S. Datta, in *Symp. VLSI Technol.* (2017), pp. 154–155.

[38] S.R. Bakaul, C.R. Serrao, M. Lee, C.W. Yeung, A. Sarker, S.L. Hsu, A.K. Yadav, L. Dedon, L. You, A.I. Khan, J.D. Clarkson, C. Hu, R. Ramesh, and S. Salahuddin, Nat. Commun. **7**, 10547 (2016).

[39] M.H. Lee, K.-T. Chen, C.-Y. Liao, S.-S. Gu, G.-Y. Siang, Y.-C. Chou, H.-Y. Chen, J. Le, R.-C. Hong, Z.-Y. Wang, S.-Y. Chen, P.-G. Chen, M. Tang, Y.-D. Lin, H.-Y. Lee, K.-S. Li, and C.W. Liu, in *IEEE Intl.*

*Electron Devices Meet.* (2018), pp. 735–738.

40 K.-S. Li, Y.-J. Wei, Y.-J. Chen, W.-C. Chiu, H.-C. Chen, M.-H. Lee, Y.-F. Chiu, F.-K. Hsueh, B.-W. Wu, P.-G. Chen, T.-Y. Lai, C.-C. Chen, J.-M. Shieh, W.-K Yeh, S. Salahuddin, and C. Hu, in *IEEE Intl. Electron Devices Meet.* (2018), pp. 731–734.

41 J. Jo and C. Shin, IEEE Electron Device Lett. **37**, 245 (2016).

42 M.H. Lee, S.-T. Fan, C.-H. Tang, P.-G. Chen, Y.-C. Chou, H.-H. Chen, J.-Y. Kuo, M.-J. Xie, S.-N. Liu, M.-H. Liao, C.-A. Jong, K.-S. Li, M.-C. Chen, and C.W. Liu, in *IEEE Intl. Electron Devices Meet.* (2016), pp. 306–309.

43 Z. Krivokapic, U. Rana, R. Galatage, A. Razavieh, A. Aziz, J. Liu, J. Shi, H.J. Kim, R. Sporer, C. Serrao, A. Busquet, P. Polakowski, J. Müller, W. Kleemeier, A. Jacob, D. Brown, A. Knorr, R. Carter, and S. Banna, in *IEEE Intl. Electron Devices Meet.* (2017), pp. 357–360.

44 M.H. Lee, P. Chen, S. Fan, Y. Chou, C. Kuo, C. Tang, H. Chen, and S. Gu, in *IEEE Intl. Electron Devices Meet.* (2017), pp. 565–568.

45 C. Fan, C. Cheng, Y. Chen, C. Liu, and C. Chang, in *IEEE Intl. Electron Devices Meet.* (2017), pp. 561–564.

46 D. Kwon, K. Chatterjee, A.J. Tan, A.K. Yadav, H. Zhou, A.B. Sachid, R. Dos Reis, C. Hu, and S. Salahuddin, IEEE Electron Device Lett. **39**, 300 (2018).

47 H. Zhou, D. Kwon, A.B. Sachid, Y. Liao, K. Chatterjee, A.J. Tan, A.K. Yadav, C. Hu, and S. Salahuddin, in *Symp. VLSI Technol.* (2018), pp. 53–54.

48 J. Zhou, G. Han, Q. Li, Y. Peng, X. Lu, C. Zhang, J. Zhang, Q.-Q. Sun, D.W. Zhang, and Y. Hao, in *IEEE Intl. Electron Devices Meet.* (2016), pp. 310–313.

49 C.-J. Su, Y.-T. Tang, Y.-C. Tsou, P.-J. Sung, F.-J. Hou, C.-J. Wang, S.-T. Chung, C.-Y. Hsieh, Y.-S. Yeh, F.-K. Hsueh, K.-H. Kao, S.-S. Chuang, C.-T. Wu, T.-Y. You, Y.-L. Jian, T.-H. Chou, Y.-L. Shen, B.-Y. Chen, G.-L. Luo, T.-C. Hong, K.-P. Huang, M.-C. Chen, Y.-J. Lee, T.-S. Chao, T.-Y. Tseng, W.-F. Wu, G.-W. Huang, J.-M. Shieh, and W.-K. Yeh, in *Symp. VLSI Technol.* (2017), pp. 152–153.

50 J. Zhou, G. Han, Y. Peng, Y. Liu, J. Zhang, Q.-Q. Sun, D.W. Zhang, and Y. Hao, IEEE Electron Device Lett. **38**, 1157 (2017).

51 W. Chung, M. Si, and P.D. Ye, in *IEEE Intl. Electron Devices Meet.* (2017), pp. 365–368.

52 C.J. Su, T.C. Hong, Y.C. Tsou, F.J. Hou, P.J. Sung, M.S. Yeh, C.C. Wan, K.H. Kao, Y.T. Tang, C.H. Chiu, C.J. Wang, S.T. Chung, T.Y. You, Y.C. Huang, C.T. Wu, K.L. Lin, G.L. Luo, K.P. Huang, Y.J. Lee, T.S. Chao, W.F. Wu, G.W. Huang, J.M. Shieh, W.K. Yeh, and Y.H. Wang, in *IEEE Intl. Electron Devices Meet.* (2017), pp. 369–372.

53 J. Zhou, J. Wu, G. Han, R. Kanyang, Y. Peng, J. Li, H. Wang, Y. Liu, J. Zhang, Q.Q. Sun, D.W. Zhang, and Y. Hao, in *IEEE Intl. Electron Devices Meet.* (2017), pp. 373–376.

54 W. Chung, M. Si, and P.D. Ye, in *2018 76th Annu. Device Res. Conf.* (IEEE, 2018).

[55] F.A. McGuire, Z. Cheng, K. Price, and A.D. Franklin, Appl. Phys. Lett. **109**, 93101 (2016).

[56] A. Nourbakhsh, A. Zubair, S.J. Joglekar, M.S. Dresselhaus, and T. Palacios, Nanoscale **9**, 6122 (2017).

[57] M. Si, C.-J. Su, C. Jiang, N.J. Conrad, H. Zhou, K.D. Maize, G. Qiu, C.-T. Wu, A. Shakouri, M.A. Alam, and P.D. Ye, Nat. Nanotechnol. **13**, 24 (2018).

[58] Z. Yu, H. Wang, W. Li, S. Xu, X. Song, S. Wang, P. Wang, P. Zhou, Y. Shi, Y. Chai, and X. Wang, in *IEEE Intl. Electron Devices Meet.* (2017), pp. 577–580.

[59] F.A. McGuire, Y.-C. Lin, K. Price, G.B. Rayner, S. Khandelwal, S. Salahuddin, and A.D. Franklin, Nano Lett. **17**, 4801 (2017).

[60] M. Si, C. Jiang, C. Su, Y. Tang, L. Yang, W. Chung, M.A. Alam, and P.D. Ye, in *IEEE Intl. Electron Devices Meet.* (2017), pp. 573–576.

[61] X. Liu, R. Liang, G. Gao, C. Pan, C. Jiang, Q. Xu, J. Luo, X. Zou, Z. Yang, L. Liao, and Z.L. Wang, Adv. Mater. **30**, 1800932 (2018).

[62] H. Zhang, Y. Chen, S. Ding, J. Wang, W. Bao, D.W. Zhang, and P. Zhou, Nanotechnology **29**, 244004 (2018).

[63] T. Srimani, G. Hills, M.D. Bishop, U. Radhakrishna, A. Zubair, R.S. Park, Y. Stein, T. Palacios, D. Antoniadis, and M.M. Shulaker, IEEE Electron Device Lett. **39**, 304 (2018).

[64] M. Si, C. Jiang, W. Chung, Y. Du, M.A. Alam, and P.D. Ye, Nano Lett. **18**, 3682 (2018).

[65] Q.H. Luc, S.H. Huynh, P. Huang, H.B. Do, M.T.H. Ha, Y.D. Jin, K.Y. Zhang, H.C. Wang, Y.K. Lin, Y.C. Lin, C. Hu, H. Iwai, and E.Y. Chang, in *Symp. VLSI Technol.* (2018), pp. 47–48.

[66] H. Ota, T. Ikegami, J. Hattori, K. Fukuda, S. Migita, A. Toriumi, and A.I. Metal, IEEE Intl. Electron Devices Meet. 318 (2016).

[67] C. Jiang, M. Si, R. Liang, J. Xu, P.D. Ye, and M.A. Alam, IEEE J. Electron Devices Soc. **6**, 189 (2018).

[68] S. Alghamdi, M. Si, L. Yang, and P.D. Ye, IEEE Int. Reliab. Phys. Symp. Proc. (2018).

[69] J.A. Kittl, B. Obradovic, D. Reddy, T. Rakshit, R.M. Hatcher, and M.S. Rodder, Appl. Phys. Lett. **113**, 042904 (2018).

[70] J. Van Houdt and P. Roussel, IEEE Electron Device Lett. **39**, 877 (2018).

[71] H. Wang, M. Yang, Q. Huang, K. Zhu, Y. Zhao, Z. Liang, C. Chen, Z. Wang, Y. Zhong, X. Zhang, and R. Huang, in *IEEE Intl. Electron Devices Meet.* (2018), pp. 707–710.

[72] C. Jin, K. Jang, T. Saraya, T. Hiramoto, and M. Kobayashi, in *IEEE Intl. Electron Devices Meet.* (2018), pp. 723–726.

[73] S. Migita, H. Ota, and A. Toriumi, in *IEEE Intl. Electron Devices Meet.* (2018), pp. 719–722.

[74] M. Jerry, J.A. Smith, K. Ni, A. Saha, S. Gupta, and S. Datta, in *2018 76th Annu. Device Res. Conf.* (2018).

[75] Z. Dong and J. Guo, IEEE Trans. Electron Devices **64**, 2927 (2017).

[76] K. Karda, C. Mouli, and M.A. Alam, IEEE Electron Device Lett. **37**, 801 (2016).

[77] W. Chung, M. Si, P.R. Shrestha, J.P. Campbell, K.P. Cheung, and P.D. Ye, in *Symp. VLSI Technol.* (2018), pp. 89–90.

[78] K. Chatterjee, A.J. Rosner, and S. Salahuddin, IEEE Electron Device Lett. **38**, 1328 (2017).

[79] D. Kwon, Y. Liao, Y. Lin, J.P. Duarte, K. Chatterjee, A.J. Tan, A.K. Yadav, C. Hu, Z. Krivokapic, and S. Salahuddin, in *Symp. VLSI Circuits Dig. Tech. Pap.* (2018), pp. 49–50.

[80] *ITRS 2.0* (2015).

[81] M.A. Alam, R.K. Smith, B.E. Weir, and P.J. Silverman, Nature **420**, 378 (2002).

[82] M.A. Alam and R.K. Smith, in *IEEE Int. Reliab. Phys. Symp.* (2003), pp. 406–411.

[83] M.A. Alam, H. Kufluoglu, D. Varghese, and S. Mahapatra, Microelectron. Reliab. **47**, 853 (2007).

[84] D. Varghese, M.A. Alam, and B. Weir, in *IEEE Int. Reliab. Phys. Symp.* (2010), pp. 1091–1094.

[85] T.P. Ma and J.P. Han, IEEE Electron Device Lett. **23**, 386 (2002).

[86] A.I. Khan, U. Radhakrishna, K. Chatterjee, S. Salahuddin, and D.A. Antoniadis, IEEE Trans. Electron Devices **63**, 4416 (2016).

[87] K. Karda and M.A. Alam, Unpublished.

[88] M. Masuduzzaman and M.A. Alam, Nano Lett. **14**, 3160 (2014).

[89] H.L. Stadler, J. Appl. Phys. **29**, 1485 (1958).

[90] J.F. Scott, L.D. McMillan, and C.A. Araujo, Ferroelectrics **93**, 31 (1989).

[91] P.K. Larsen, G.L.M. Kampschöer, M.J.E. Ulenaers, G.A.C.M. Spierings, and R. Cuppens, Appl. Phys. Lett. **59**, 611 (1991).

[92] P.K. Larsen, G.L.M. Kampschoer, M.B. van der Mark, and M. Klee, in *Proc. Eighth IEEE Int. Symp. Appl. Ferroelectr.* (1992), pp. 217–224.

[93] J. Li, B. Nagaraj, H. Liang, W. Cao, C.H. Lee, and R. Ramesh, Appl. Phys. Lett. **84**, 1174 (2004).

[94] H. Ishii, T. Nakajima, Y. Takahashi, and T. Furukawa, Appl. Phys. Express **4**, (2011).

[95] J. Müller, T.S. Böscke, U. Schröder, R. Hoffmann, T. Mikolajick, and L. Frey, IEEE Electron Device Lett. **33**, 185 (2012).

[96] E. Yurchuk, J. Müller, J. Paul, T. Schlösser, D. Martin, R. Hoffmann, S. Müeller, S. Slesazeck, U.S. Member, R. Boschke, R. Van Bentum, T. Mikolajick, and S. Member, IEEE Trans. Electron

Devices **61**, 3699 (2014).

[97] S. Dünkel, M. Trentzsch, R. Richter, P. Moll, C. Fuchs, O. Gehring, M. Majer, S. Wittek, B. Müller, T. Melde, H. Mulaosmanovic, S. Slesazeck, S. Müller, J. Ocker, M. Noack, D.A. Löhr, P. Polakowski, J. Müller, T. Mikolajick, J. Höntschel, B. Rice, J. Pellerin, and S. Beyer, in *IEEE Intl. Electron Devices Meet.* (2017), pp. 485–488.

[98] H. Mulaos, J.O. Ker, S. Mulle, U. Schroeder, J. Müller, P. Polakowski, S. Flachowsky, R. Van Bentum, T. Mikolajick, and S. Slesazeck, ACS Appl. Mater. Interfaces **9**, 3792 (2017).

[99] K. Karda, A. Jain, C. Mouli, M.A. Alam, K. Karda, A. Jain, C. Mouli, and M.A. Alam, Appl. Phys. Lett. **106**, 163501 (2015).

[100] A. Jain and M.A. Alam, IEEE Trans. Electron Devices **60**, 4269 (2013).

**Supplementary Materials**

**An Overview of the Theory of Phase Transition**

NC-FET results have been interpreted by several different modeling approaches. The difference in the modeling approaches and the parameterization needed to fit the experiments sometimes lead to conflicting interpretations. The key challenge is to understand the role of domain dynamics. The exact problem is unsolved and potentially unsolvable, but a summary of the modeling approaches may be helpful.

The time dynamics of ferroelectric capacitor is often described by Ginzburg-Landau theory

$$M_0 \, d^2P/dt^2 + L_0 \, dP/dt = - \, dU/dP \qquad (1a)$$

where P is the net polarization, $M_0$ is the effective mass of the atoms being displaced, $L_0$ is phonon-induced "friction" and

$$U(P, V_{FE}, k) = U_{FE}(P, V_{FE}) + k \, (\nabla P)^2 \qquad (1b)$$

The key debate concerns the form of $U_{FE}(P, V_{FE})$ and if the domain dynamics can be used to renormalize M, L, and $U_{FE}$. Here, k is the domain coupling constant, which is related to spatial non-uniformity of P. We note that by definition:

$$\frac{dU_{FE}}{dQ} = V_{FE}, \text{or } C_F^{-1} = \frac{d^2(U_{FE})}{dQ^2}, \text{and } C_{FE} = \frac{dP}{dV_{FE}}$$

So that Eq. (1a) and (1b) can be written in variety of equivalent forms. The equation is solved subject to the boundary conditions of interfacial stress and charge/polarization continuity.

Four types of approximations have been used so far for solving the Ginzburg-Landau equations:

**1. Classical Landau-Khalatnokov (L-K) Model based on Renormalized Domain Dynamics:**
In this formulation, Eq. (1b) is first written as

$$U_{FE}(P, V_{FE}) = \alpha^* \, P^2 + \beta^* \, P^4 + \gamma^* \, P^6 - P \, V_{FE} + k \, (\nabla P)^2$$
$$\rightarrow \alpha P^2 + \beta \, P^4 + \gamma \, P^6 - P \, V_{FE}. \qquad (2a)$$

Here the polarization anisotropy factors are renormalized by including the $k(\nabla P)^2$ term either by using a two-domain approximation, or by fitting the experimental data. The empirical approach can also be used to renormalize M and L parameters by fitting time-dependent Fe-FET P-E data (by indirectly including the effect of k ). The key to this analysis is that $\alpha < 0$ following the renormalization. Most transistor analysis (steady state and transient) are based on the renormalized L-K equation. In particular, the following equation is used to interpret the results in steady state

$$\frac{dU_{FE}}{dP} = 0 = 2\alpha \, P + 4 \, \beta \, P^3 + 5 \, \gamma \, P^5 - V_{FE}. \qquad (2b)$$

The original L-K model suggests the second derivative of $U_{FE}$ is positive for stable physical system. The NC-FET concept suggests a linear capacitor can stabilize the FE unstable NC state with the second derivative of $U_{FE}$ negative.[5]

2. **KAI Model based on Explicit Time Approximation**: The second approach, proposed and published by Y.J. Kim et al.[19], involves assuming

$$M_0 = 0, U_{FE}(P, V_{FE}) = \alpha(k, V_{FE})P - P V_{FE}, \text{ and } L_0 \to L(V, k) \qquad (3a)$$

so that the effect of domain dynamics incorporated indirectly in $\alpha$ and $L$. Most important, here $\alpha > 0$. With these approximations, the resulting first order equation (i.e. $L\, dP/dt = \alpha (P_0 - P)$ is directly solvable. If $\alpha/L \equiv \left(\frac{t}{\tau}\right)^{\beta-1}$ in analogy to Avrami equation for crystal nucleation and growth, the solution of the corresponding equation

$$d\left(\frac{P}{P_0}\right)/dt = (t/\tau(V_{FE}, k))^{\beta-1} (1 - P/P_0)$$

yields the Weibull form

$$P(t) = P_0 \left(1 - \exp\left(-\left(\frac{t}{\tau}\right)^{\beta}\right)\right). \qquad (3b)$$

Merz's law makes the voltage-dependence explicit: $\tau = \tau_0 \exp(\alpha/V_{FE})$. The most important point is $\alpha > 0$ and there is no NC effect.

3. **Preisach-Miller model involving Implicit Time Dependent Approximation** [20,22]:
In a slightly different approximation, one assumes that the steady state relationship between $V_{FE}$ and $P$ is given not by Eq. (2a), but rather by two branches of the P-V relationship (with positive $\alpha$)

$$\frac{dU_{FE}}{dP} = 2\alpha P - V_{FE}, \text{ where } P \equiv P_s \left[\tanh\left(w\left(V_{FE} \pm V_c\right) + \frac{\eta V_{FE}}{V_p}\right)\right]$$

$$\text{and } w = \sigma/(2V_c) \ln (P_S + P_R)/(P_S - P_R)$$

Here, $P_S$ is the saturation polarization, $P_R$ is the remnant polarization, $V_c$ is the coercive voltage, $V_p$ is the peak voltage for $V_{FE}$. Here, $\eta$ and $\sigma$ captures $V_c$ distribution of FE domains and non-ferroelectric linear response of P.

$$L \frac{dP}{dt} = V_{FE} - 2\alpha P(t)$$

Or rewriting in the equivalent circuit format, with $V_A \to V_{FE}$, $C_{FE} = \frac{dP}{dV_{FE}}$, $\tau \equiv RC_{FE} \to LC_{FE}$, we find

$$\tau\, dV_{FE}/dt = V_A - V_{FE} \to L \frac{dP}{dt} = V_A - C_{FE}^{-1} P = V_A - 2\alpha P$$

$$C_{FE} = \frac{dP}{dV_{FE}} = (2\alpha)^{-1}$$

4. **Simplified Miller Model**: Finally, a quasi-static variant of the Miller Model (with $M_0 = 0, L_0 = 0$) is used in Ref. [70] where $\left(\frac{d^2U_{FE}}{dP^2}\right)^{-1} = C_{FE} = C_0 + \beta V_{FE}^n$. Parameter $\beta$ and $n$ are adjusted so that a double integration of the capacitance function would produce the Miller equation.

**Interpretation of Experiments.**

**1. Time-dependent Analysis:**

KAI, Miller, and modified Miller Models have been used to argue that the phase lag between applied voltage and domain flipping can also interpret the results reported in refs. [19–24,69,70]. The authors explain several features of the time-dependent data without invoking negative capacitance (i.e. $\alpha < 0, \beta < 0$). The model anticipates response time limited by domain dynamics and could be as slow as tens of MHz range.

Time-dependence can also be interpreted by the renormalized L-K model, where the domain dynamics with empirical mass and alpha parameters are applied. Once again, the response obtained is in the tens of MHz range.

A recent experiment shows optical evidence of much faster response[78] and direct pulse measurement with a single pulse switch time of 3.6 ns and a train of pulses with pulse width down to 100 ps[77]. The direct pulse response is limited by many parasitic effects of the measured device and the ultra-fast pulse might not be able to provide sufficient polarization charges for a full ferroelectric switch through inversion charges at a single ultra-short pulse time. It is not clear if these results can or should be interpreted by the generalized Landau framework.

**2. DC Analysis:**

The renormalized L-K model has been used to interpret the subthreshold slope, negative DIBL, and NDR reported by various groups[48,53,57]. Both 3D simulation of advanced transistors[66] and compact model for 3D and 2D transistors[67] have been proposed. The model has also been used to understand the scaling consequences (e.g. inclusion of quantum and parasitic capacitance, etc.) as well as new device design integrating various "Landau" devices (FE, AFE, MEMS, piezo)[99,100]. The reduction of noise with increasing dielectric thickness[68] is another example of such an analysis producing self-consistent results.

The KAI, Miller, and/or Modified Miller models interpret the steady state response by suggesting that all steady-state measurements are in-fact time-dependent, defined by the rate of measurement. Thus, they interpret the "DC" subthreshold slope as a consequence of time-dependent phase-lag.